# Validated two-dimensional modeling of short carbon arcs: anode and cathode spots

J. Chen[1, 2,†], A. Khrabry[1,*], I. D. Kaganovich[1], A. Khodak[1], V. Vekselman[1], H.-P. Li[2,†]

[1]*Princeton Plasma Physics Laboratory, Princeton NJ 08542, USA*
[2]*Department of Engineering Physics, Tsinghua University, Beijing 100084, China*

**Abstract:**

In order to study properties of short carbon arcs, a self-consistent model was implemented into a CFD code ANSYS-CFX. The model treats transport of heat and electric current in the plasma and the electrodes in a coupled manner and accounts for gas convection in the chamber. Multiple surface processes at the electrodes are modeled, including the formation of space-charge limited sheaths, ablation and deposition of carbon, emission and absorption of radiation and electrons. The simulations show that the arc is constricted near the cathode and the anode front surfaces leading to the formation of electrode spots. The cathode spot is a well-known phenomenon and mechanisms of its formation were reported elsewhere. However, the anode spot formation mechanism discovered in this work was not reported before. We conclude that the spot formation is not related to plasma instability, as commonly believed in case of constricted discharge columns, but rather occurs due to the highly nonlinear nature of heat balance in the anode. We additionally demonstrate this property with a reduced anode heat transfer model. We also show that the spot size increases with the arc current. This anode spot behavior was also confirmed in our experiments. Due to the anode spot formation, a large gradient of carbon gas density occurs near the anode, which drives a portion of the ablated carbon back to the anode at its periphery. This can consequently reduce the total ablation rate. Simulation results also show that the arc can reach local chemical equilibrium (LCE) state in the column region while the local thermal equilibrium (LTE) state is not typically achieved for experimental conditions. It shows that it is important to account for different electron and gas temperatures in the modeling of short carbon arcs.

## I. INTRODUCTION

Over the past few decades, the production of carbon nanoparticles has attracted a lot of attention, and many approaches, such as chemical vapor deposition[1], plasma-enhanced chemical vapor deposition[2], laser vaporization[3] and arc discharge[4-7], are utilized for the synthesis. Among

---

these methods, an atmospheric pressure arc discharge is the simplest and cheapest method that meets the requirements of the industrial-scale production[8]. In carbon arc discharges, a hot highly-ablating anode is usually used as a source of material for the growth of nanoparticles. The typical surface temperature of the anode tip can reach ~4000 K. At such temperature carbon ablation, strong thermionic electron emission and high radiation are all significant. The system of equations describing the entire arc is highly nonlinear and governed by the coupling of particles and heat transport in both plasma and electrodes.

One of the interesting features of the carbon arc is the formation of the electrode spots, which have already been observed in many carbon arc experiments [7, 9]. Studies regarding the spots in glow and arc discharges are conducted by many groups including both the numerical simulations [10-13] and experiments[14,15]. Such spots may significantly affect material processing in electric arcs. Reference [14] claims that arc constriction and formation of anode spots occur due to overheating instability. In Reference [15], it was observed that anode spots disappear with an increase of the electrode electrical conductivity. Recent numerical studies regarding electrode spots are reviewed in Refs. [16] and [17]. It is concluded that self-organization and bifurcation theories can be applied to explain the spots. It was also proposed that this phenomenon could be related to the plasma instability, e.g., the thermal instability[11, 12]. However, to our knowledge, formation of anode spots in carbon arcs has not been analyzed until now. Both the underlying mechanisms, and the effects of the electrode spots on the nanoparticle's synthesis are still not clear. Understanding the formation of the spots and their effect on the heat transfer, ablation, deposition and transport of carbon was the motivation for the present study.

To correctly model the two-dimensional arc profiles, we developed a self-consistent arc model. So far, many numerical models have been established to simulate arc discharges[18-31]. Some of them are developed to describe only specific regions [18, 19, 20, 23] and some are for the entire arc[21, 22, 24-31]. Reviews of the plasma-electrode interaction in arc discharges and argon arc modeling can be found in Ref. 32 and Ref. 33. To simplify the modeling, assumptions of local equilibriums, e.g., local thermal equilibrium (LTE) and local chemical equilibrium (LCE) are often made. If LTE is assumed, there is no need to solve the electron energy equation but regard the electron temperature the same as that of heavy particles; the LCE approximation allows us to use the Saha equation instead of multi-species transport equations to obtain species number densities. These assumptions provide a great convenience by reducing the computational difficulties. However, these models do not take into account non-equilibrium effects and introduce artificial features in plasma profiles, especially in the near-electrode regions[26,27]. For typical carbon arc discharges, arc length is very short (usually several millimeters), and the size of non-equilibrium regions becomes comparable to the arc column length. Hence, non-equilibrium effects are important. For example, as shown in Ref. 27, the ion current and plasma density near the electrodes would be drastically lower if non-equilibrium effects near the cathode are neglected.

Therefore, in this paper to study the characteristics of short carbon arc discharges, we developed a self-consistent arc model without any equilibrium assumptions. The entire arc is modeled including the cathode, anode, arc column and gas flow in a large chamber region surrounding the arc. Sheath is not resolved but taken into account as boundary conditions. Ablation/deposition, radiation and electron emission are all taken into consideration.

This paper is organized as follows. Section II describes the details of the self-consistent arc model. Section III describes the spot formation in short arc discharges. Section IV provides a summary. Appendixes A-E gives additional details of the model.

## II. SELF-CONSISTENT ARC MODEL

In this section, the system of equations self-consistently describing the arc is thoroughly introduced. Here we only focus on the arc and not on nanoparticle synthesis, see e.g. Ref.[4], which is left for later modeling. Therefore, only the carbon and helium atoms, electrons and singly charged carbon ions are considered. Helium is assumed to be not ionized due to the large ionization potential. Also, for the fast convergence of the simulations, we only perform the steady-state modeling and leave time-dependent modeling for future work.

### A. Governing equations
#### 1. Continuity and momentum equations for carbon and helium atoms

The steady-state continuity and momentum conservation equations for the carbon-helium gas mixture (i.e. the Navier-Stokes equations) can be expressed in the following form:

$$\nabla \cdot (\rho \vec{v}) = 0, \quad (1)$$

$$\nabla \cdot (\rho \vec{v}\vec{v}) = -\nabla p + \nabla \cdot \tau + \rho \vec{g}, \quad (2)$$

where $\rho = \sum_s n_s m_s$ is the mixture density, where $n_s$ and $m_s$ are the number density and mass of species $s$, $s$ represents all the species including electrons, carbon ions and atoms, and helium atoms, $\vec{v}$ is the flow velocity, $p$ is the pressure, $\tau$ is the viscosity stress tensor and $\vec{g}$ is the gravity acceleration. The viscosity of the C-He mixture depends on the temperature, the mixing ratio and background pressure. However, for the temperature range of our interest (<10000 K), the viscosity depends weakly on the pressure and mixing ratios[34], and, therefore, is replaced by the viscosity of the pure helium at 1 atm (still accounting for the temperature dependence).

Equation of state is

$$p = (n_C + n_{He} + n_i)kT + n_e kT_e, \quad (3)$$

where $T_e$ is the electron temperature and $T$ is the temperature of heavy species (note that all heavy particles have the same temperature).

#### 2. Transport equations for the mixture species
Carbon gas transport equation is

$$\nabla \cdot (\rho c_C \vec{v}) = \nabla \left[ D_{C, He} \nabla (\rho c_C) \right] - m_C S_i, \quad (4)$$

where $c_C$ is the mass fraction of carbon atoms in the mixture, $D_{C, He}$ is the diffusion coefficient for carbon in helium, $m_C$ is mass of carbon atom, and $S_i$ is the net ionization rate of carbon atoms. $D_{C,}$

$_{He}$ was calculated using Eq. (5)[27], with the momentum transfer cross-sections $\sigma_{C,He}=3\times10^{-19}$ m$^2$ taken from Ref. 35:

$$D_{C,He} = \frac{3\pi}{32}\sqrt{\frac{8kT}{\pi m_{C,He}}} \frac{1}{(n_C+n_{He})\sigma_{C,He}}, \quad m_{C,He} = \frac{m_C m_{He}}{m_C + m_{He}}. \tag{5}$$

The net ionization rate can be written as

$$S_i = k_i n_e n_C - k_r n_e^3, \tag{6}$$

where $k_i$, $k_r$ are the step-wise ionization and recombination rate coefficients for carbon atoms taken from Ref. 36.

Note that here quasi-neutrality condition is used for the entire arc:

$$n_e = n_i,$$

and the space-charge sheaths near the electrodes are treated as boundary conditions, described in Sec. II (b).

The excitation energy and direct ionization energy for helium atoms are respectively 19.8 eV and 24.6 eV[28], which are considerably higher than the ionization energy for carbon atoms, 11.26 eV, therefore, helium excited atoms and ions can be omitted from consideration. For the same reason, multi-charged carbon ions are also omitted, and the transport equation for carbon ion, C$^+$ reads

$$\nabla \cdot \frac{\vec{j}_i}{e} = S_i, \tag{7}$$

where $\vec{j}_i$ is the ion current. Electron and ion currents are given by

$$\vec{j}_e = \sigma \frac{k}{e}\nabla T_e + \sigma \frac{k}{e}T_e \nabla \ln n_e + \sigma \vec{E} - en_e \vec{v}, \quad \vec{j}_i = -\mu_i k n_i \nabla T - \mu_i kT \nabla n_i + e\mu_i n_i E + en_i \vec{v},$$

$$\sigma = en_e \mu_e = \frac{n_e e^2}{m_e(\nu_{e,a}+\nu_{e,i})}, \quad \mu_i = \frac{e}{m_i(\nu_{i,a}+\nu_{i,e})} \tag{8}$$

where $\sigma$ is the electrical conductivity, $\vec{j}_e$, $\vec{j}_i$, and $\vec{E}$ are the electron current density, the ion current density and the electric field, $\mu_e$ and $\mu_i$ are the mobilities of electrons and ions. Note that $\vec{v}$ is the flow velocity of the carbon-helium mixture but not the individual species velocity. $\nu_{k,j}$ is the effective collision frequency of $k$ species with $j$ species

$$\nu_{k,j} = \frac{4}{3}\sqrt{\frac{8kT_{kj}}{\pi m_{kj}}} C_{kj}\sigma_{kj} n_j, \quad T_{kj} = \frac{m_k T_j + m_j T_k}{m_k + m_j}, \quad m_{kj} = \frac{m_k m_j}{m_k + m_j}, \tag{9}$$

where $T_{kj}$, $m_{kj}$ and $\sigma_{kj}$ are the binary temperatures, binary masses and momentum transfer cross-sections. $C_{kj}$ are the coefficients derived from kinetic theory[27]. The Coulomb collisions between electrons and ions are described by the following cross-section:

$$\sigma_{ei} = \frac{e^4 \ln\Lambda}{32\pi\varepsilon_0^2 (kT_e)^2}, \tag{10}$$

where $\varepsilon_0$ is the vacuum permittivity, $k$ is the Boltzmann constant, $T_e$ is the electron temperature and $\ln\Lambda$ is the Coulomb logarithm. In this work, Eq. (7) was implemented into ANSYS-CFX in the ambipolar form (see Appendix A for details).

The electric field can be calculated from Eq. (8) for the electron current,

$$\vec{E} = -\nabla V = -\frac{k}{e}\nabla T_e - \frac{k}{e}T_e \nabla \ln n_e + \frac{\vec{j}_e}{\sigma}, \tag{11}$$

Where the last term can be neglected for most of the conditions in the short arc, where the terms with gradients are dominant, see detailed analysis in Ref. [49].

Omitting the radiative loss from plasma (see Appendix B), electron and ion energy balance equations are

$$\nabla \cdot \left(2.5kT_e \frac{\vec{j}_e}{e}\right) = \nabla \cdot (\lambda_e \nabla T_e) - S_i E_{ion} - Q^{e-h} + \vec{j}_e \cdot \vec{E}, \tag{12}$$

$$\nabla \cdot (\rho h \vec{v}) = \nabla \cdot (\lambda_h \nabla T) + \nabla \cdot (\vec{v} \cdot \tau) + Q^{e-h} + \vec{j}_i \cdot \vec{E}, \tag{13}$$

where $\lambda_e$ and $\lambda_h$ are the thermal conductivities for electrons and heavy particles. $\lambda_e$ is obtained from Ref. 37, and $\lambda_h$ is from Ref. 34. $E_{ion}$ is the direct ionization energy of carbon atoms. $h$ is the total enthalpy computed as $h = \int c_{p,mix}(T) dT + \frac{1}{2}v^2$, $c_{p,mix} = c_{p,He} \cdot (1 - c_C) + c_{p,C} \cdot c_C$. The specific heat for helium, $c_{p,He}$, is constant (5240 J/kg/K) and for carbon (including carbon atoms and ions), $c_{p,C}$, is taken from Ref. 39. $Q^{e-h}$ represents the heat exchange between electrons and heavy particles due to elastic collisions, following the form[28]

$$Q^{e-h} = n_e \sum_s \frac{3}{2} k(T_e - T) \frac{2m_e}{m_s} \nu_{e,s}. \tag{14}$$

### 3. Conservation of electric current used to determine the profile of electric potential

The current conservation equation is

$$\nabla \cdot \vec{j} = 0. \tag{15}$$

Here, $\vec{j} = \vec{j}_e + \vec{j}_i$ is the total current density. Substituting Eq. (11) into Eq. (15), and neglecting the divergence of ion current density[11, 20] we obtain an elliptic equation for the electric potential

$$\nabla \cdot (\sigma \nabla V) = \nabla \cdot \left(\sigma \frac{k}{e}\nabla T_e\right) + \nabla \cdot \left(\sigma \frac{k}{e}T_e \nabla \ln n_e\right). \tag{16}$$

This equation is supplemented with the current conservation equation in electrodes,

$$\nabla \cdot \vec{j} = \nabla \cdot (\sigma_{graphite} \nabla V) = 0.$$

In addition, the heat conduction equation with Joule heating is solved inside the cathode and anode. The thermal conductivity, $\lambda_{graphite}$, and the electrical conductivity of graphite, $\sigma_{graphite}$, are taken as functions of temperature according to Ref. 40.

All the above equations are implemented into a computational framework ANSYS-CFX. ANSYS-CFX is a robust Computational Fluid Dynamics (CFD) code designed for 3D simulations in complex geometry using unstructured grids and multiple processors. The code incorporates models for various physical phenomena, including the heat and mass transfer, radiation transfer, buoyancy effects, etc. The code is extendable and enables the implementation of custom-based processes by introducing additional variables and transport equations with the custom-based boundary conditions. This feature of the code was utilized for the implementation of the self-consistent arc model, for the gas/plasma and solid (electrode) domains.

## B. Computational domain and boundary conditions

In this paper, an axisymmetric computational domain is adopted we set the origin at the center of cathode tip surface), see Fig. 1. The arc model is designed to describe the experiments reported in Refs. [4, 41]. The entire arc setup, including the graphite cathode, graphite anode and surrounding chamber filled with the background gas are included in the model. The cathode and anode are located along the axis of the chamber. The chamber radius and height are 78 mm and 240 mm, respectively. The length of the anode rod is 126 mm. The length of the cathode rod was different for different inter-electrode gap widths, e.g., 111 mm for a gap width $d=3$ mm. Radii of the cathode and the anode are 4.75 mm and 3.2 mm, respectively. A small outlet is placed at the bottom of the chamber wall in order to maintain the background pressure, similar to the experiments. The width of the outlet is 5 mm. In the simulations, the inter-electrode gap width is fixed. A constant gap width was also maintained in actual experiments by automatically adjusting the positions of the electrodes as the anode material was evaporated by the arc and the deposit grew on the cathode.

Boundary conditions for all the equations are listed in Table 1. The no-slip assumption, i.e. zero gas velocity, is adopted for all solid walls, and the outlet is defined as the open boundary where free flow in and out is allowed with a constant pressure 68 kPa. Boundary conditions strongly affect plasma and gas properties: the current density and heat flux are affected by the sheath boundary conditions, and the carbon density by carbon ablation of and deposition on electrodes.

Table I Boundary conditions

| Equations | Continuity & momentum[*] | Carbon gas transport[*] | Poisson Equation | Ion transport | Electron energy[@@] | Heavy particle and electrodes energy[@@] |
|---|---|---|---|---|---|---|
| Cathode surface | $\vec{n}\cdot\vec{\Gamma}=\Gamma_{abl/depos}$ | $\vec{n}\cdot\vec{\Gamma}_C=\Gamma_{abl/depos}$ | $V_{electrode}=V_{plasma}+V_{sheath}$ $\vec{n}\cdot\sigma\nabla V_{electrode}=\vec{n}\cdot\vec{j}$ ** | Eq. (20)/(21)[@] | $\vec{n}\cdot\lambda_e\nabla T_e=q_e$ | $T_{electrode}=T$ $\vec{n}\cdot\lambda_h\nabla T - \vec{n}\cdot\lambda_{graphite}\nabla T_{electrode}=q_h$ |
| Anode surface | $\vec{n}\cdot\vec{\Gamma}=\Gamma_{abl/depos}$ | $\vec{n}\cdot\vec{\Gamma}_C=\Gamma_{abl/depos}$ | $V_{electrode}=V_{plasma}+V_{sheath}$ ** $\vec{n}\cdot\sigma\nabla V_{electrode}=\vec{n}\cdot\vec{j}$ | Eq.(20)/(21)[@] | $\vec{n}\cdot\lambda_e\nabla T_e=q_e$ | $T_{electrode}=T$ $\vec{n}\cdot\lambda_h\nabla T - \vec{n}\cdot\lambda_{graphite}\nabla T_{electrode}=q_h$ |
| Chamber surface | $\vec{n}\cdot\vec{\Gamma}=\Gamma_{abl/depos}$ | $\vec{n}\cdot\vec{\Gamma}_C=\Gamma_{abl/depos}$ | $\vec{n}\cdot\sigma\nabla V=0$ | $\vec{n}\cdot\vec{j}_i=0$ | $T_e$=350 K | $T$=350 K |
| Outlet | $p$=68 kPa | $c_C$=0 | $\vec{n}\cdot\sigma\nabla V=0$ | $\vec{n}\cdot\vec{j}_i=0$ | $T_e$=350 K | $T$=350 K |
| Cathode outer tip | - | - | $V$=0 | - | - | $T$=350 K |
| Anode outer tip | - | - | $V=V_{out}$*** | - | - | $T$=350 K |

[*]Tangential velocity is zero (non-slip condition)

**Sheath voltage drop, $V_{sheath}$, is calculated according to the local current balance at the plasma-solid interface [Eq. (19)]. See details in Ref. [27]

***$V_{out}$ is calculated in the solver iterations by satisfying that the actual total current equals to the given arc current

@For $V_{sheath}>0$ (and $V_{sheath}<0$), ion fluxes are determined from Eq. (20) [and Eq. (21)]

@@$q_e$, $q_h$ are given by Eq. (25) and Eq. (26)

In Table I, $\vec{n}$, $\vec{\Gamma}$ and $\vec{\Gamma}_C$ represent the unit vector normal to the surface, the total flux of the carbon/helium gas mixture, and the carbon gas flux, respectively. To account for the ablation and deposition processes, a new relation for the boundary mass fluxes, $\Gamma_{abl/depos}$, is used to enable the automatic selection between the ablation and deposition

$$\Gamma_{abl/depos} = \left[ p_{sat,C}(T) - p_C \right] \sqrt{\frac{M_C}{2\pi k T}}. \tag{17}$$

Here, $p_C = n_C k T$ is the partial pressure of carbon gas at the electrode surface, and $p_{sat,C}(T)$ is the saturated pressure at the electrode surface as a function of gas temperature, $T$, which can be determined from the Clapeyron-Clausius relation:

$$p_{sat,C}(T) = 1\,[\text{atm}] \cdot \exp\left[ -\frac{q_C M_C}{k_B} \left( \frac{1}{T} - \frac{1}{T_{sat,1atm}} \right) \right], \tag{18}$$

where $T_{sat,1atm}$=3900 K is the saturated temperature of carbon vapor at 1 atm. $q_C$=5.93×10$^7$ J/kg is the latent heat for atomic carbon [42]. Note that Eq. (17) includes the effect of carbon gas pressure which was missed in the previous modelling[29]. Combining Eq. (17) as the boundary condition, the carbon gas transport with the electrode ablation and deposition can be accurately solved using Eq. (4).

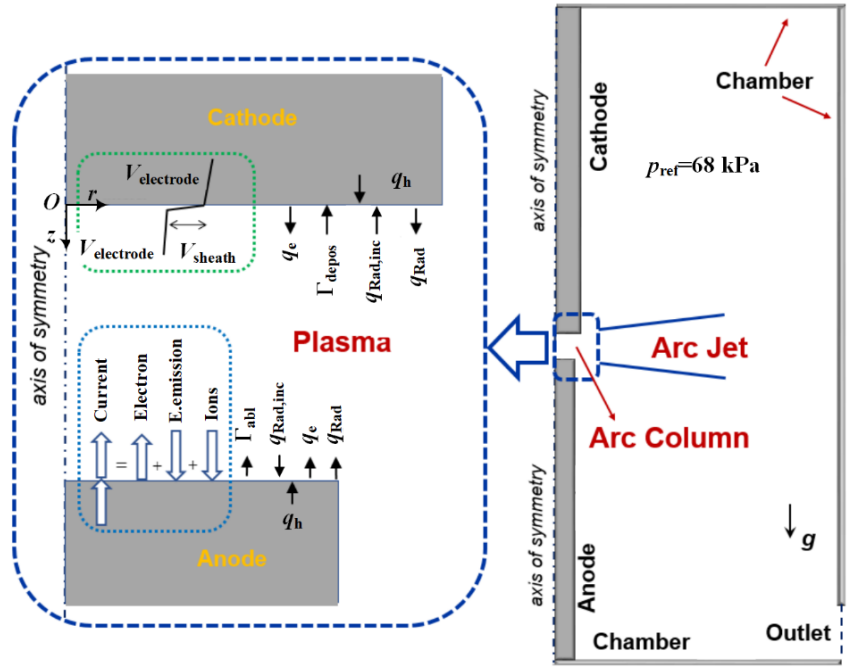

**Fig. 1.** Schematic of simulations set up, left: showing the boundary conditions at the electrodes, right: full simulation domain.

In our model, sheath width is assumed very small (typically of the order of micrometer) and is not resolved. Instead, sheath is described with the effective boundary conditions for the equations for potential and energy balance. The adopted sheath model has been already benchmarked in Ref. 27.

Current balance equation at the electrode surfaces reads

$$\vec{j} = \vec{j}_i + \vec{j}_e = \vec{j}_i + \vec{j}_e^{\text{emission}} + \vec{j}_e^{\text{plasma}}, \tag{19}$$

where $\vec{j}$, $\vec{j}_e^{\text{emission}}$ and $\vec{j}_e^{\text{plasma}}$ are the total current density, emission current density and plasma bulk electron current density at the boundary, respectively. In the case of positive sheath voltage (the wall potential is positive relative to the plasma potential), the boundary ion current is written as

$$\vec{n} \cdot \vec{j}_i = -\frac{1}{4} e n_i \sqrt{\frac{8kT}{\pi M_C}} \exp\left(-\frac{eV_{\text{sheath}}}{kT}\right). \tag{20}$$

For the negative sheath voltage, the ion current is determined by the Bohm's criterion

$$\vec{n} \cdot \vec{j}_i = -e n_i \sqrt{\frac{k(T_e + T)}{M_C}}. \tag{21}$$

Eq. (20) and Eq. (21) give the boundary conditions at the cathode and anode surface for ion transport equation (see Table I). At the chamber wall and outlet far from the arc bulk, the ion density is negligible and assumed equal to zero.

The electron current from the plasma to the wall is given by

$$\vec{n} \cdot \vec{j}_e^{\text{plasma}} = \frac{1}{4} e n_e \sqrt{\frac{8kT_e}{\pi m_e}} \exp\left(\frac{e \min(V_{\text{sheath}}, 0)}{kT_e}\right). \tag{22}$$

Note that the electron emission is very important and affects the sheath voltage drop[27, 43, 44]. Due to the high surface temperature, the thermionic emission current from the wall to the plasma should be included and is given by

$$\vec{n} \cdot \vec{j}_e^{\text{emission}} = -j_R \exp\left(\frac{-e \max(V_{\text{sheath}}, 0)}{kT_{\text{electrode}}}\right), \tag{23}$$

where $j_R$ is the Richardson's emission current

$$j_R = A_R T_{\text{electrode}}^2 \exp\left(-\frac{e(V_w + E_{\text{Schott}})}{kT_{\text{electrode}}}\right), \tag{24}$$

where $A_R$ is the Richardson's constant, $V_w$ is the work function of the electrode material (for graphite, $V_w$=4.6 V), and $E_{\text{Schott}}$=0.1 V is the Schottky correction voltage[18].

The electron heat flux at the boundary is

$$q_e = -\left(\frac{2.5kT_{\text{electrode}}}{e} - \min(V_{\text{sheath}}, 0)\right) \vec{n} \cdot \vec{j}_e^{\text{emission}} - \left(\frac{2.5kT_e}{e} - \min(V_{\text{sheath}}, 0)\right) \vec{n} \cdot \vec{j}_e^{\text{plasma}} + \frac{2.5kT}{e} \vec{j}_e, \tag{25}$$

and the boundary heat flux for the carbon-helium mixture coupled to an electrode is

$$q_h = \left(\frac{2.5kT_{electrode}}{e} + V_w + \max(V_{sheath}, 0)\right)\vec{n}\cdot\vec{j}_e^{emission} + \left(\frac{2.5kT_e}{e} + eV_w + \max(V_{sheath}, 0)\right)\vec{n}\cdot\vec{j}_e^{plasma}, \quad (26)$$

$$-\vec{n}\cdot\vec{j}_i^{plasma}\left(E_{ion} - V_w - \max(V_{sheath}, 0)\right) - q_{C,vapor}\Gamma_{abl/depos} - \varepsilon\sigma_{SB}T^4 + \varepsilon q_{Rad,inc}$$

where $q_{C,vapor}$ is the latent heat for vaporization, $\varepsilon$ is the emissivity, $\sigma_{SB}$ is the Stephen-Boltzmann coefficient and $q_{Rad,inc}$ is the incident radiation. The emissivity of the two electrodes is set as 0.8, which is a typical quantity for graphite[5]; and plasma and gas are assumed to be a transparent media, i.e. radiation and absorption from it was neglected. This is also confirmed in our experimental measurements. Monto-Carlo method is adopted for calculating the incident radiation. The model is already implemented in ANSYS; it treats all emissive walls as photon sources and tracks photons generated by these sources until they are absorbed. Each time the surface experiences an intersection with a photon, the incident radiation of interest gets updated (see ANSYS CFX-Solver Modeling Guide[38] for more details).

It should be emphasized that the present sheath model enables one to find a self-consistent solution that takes into account the sheath voltage drops, boundary current and heat flux without any prior assumption and allows for a solution of a highly nonlinear system of equations describing an arc. Non-uniform grids were employed for better resolving the non-equilibrium transport near the electrodes. Convergency check of grid resolution was also performed and is described in Appendix C.

## III. RESULTS

### A. Validation against experimental data

For validating the CFX code, two series of simulations with the experimental conditions of Ref. 41 were performed. In the experiments, the background pressure is 67 kPa. Plasma density and electron temperature were measured using the combination of optical emission spectroscopy, spectrally-resolved fast frame imaging and planar laser induced fluorescence. Ablation and deposition rates were measured by electrodes weighting after arc run of 60 s [41]. Note that no a priori equilibrium assumption was taken in the measurements. In low ablation regime, the measured electron temperature in the arc core for the case with $d$=3.0 mm and $I$=50 A is 0.8 ± 0.1 eV in agreement with the simulated value, 0.83 eV for the same arc parameters. Simulation results also show that the arc characteristics depend on heat flux through the graphite electrodes. However, the thermal conductivity of graphite used in the electrodes can vary significantly. Therefore, we performed simulations with two values of the electrode thermal conductivity, $\lambda_{graphite}$, one was taken from Ref. [40] and the other use twice large value of $\lambda_{graphite}$. A comparison of the simulated anode ablation rate and cathode deposition rate against experimental measurement data is shown in Fig. 2(a). Inter-electrode gap width was fixed at 1.5 mm and the total arc current was varied. Qualitatively, simulated ablation and deposition rates agree with the measured ones, but the quantitative difference exists, especially for the cases with high arc current. One of the reasons is probably owing to the limitation of steady-state simulation. In reality, the arc is not exactly steady but keeps moving (usually more wildly with higher arc current). In addition, the uncertainties of anode properties in the experiments may introduce some discrepancies as well. The big difference between the results of two simulations emphasizes the significant effect of the anode thermal properties on the ablation and deposition rates. Until now, at least to our knowledge, anode properties are scarcely measured in actual experiments, and their effects on arc plasma seem not to be considered rigorously. This calls for further experimental studies to be conducted on this effect.

Radial profiles of plasma density at the mid-plane obtained by simulations and the measurement data are plotted in Fig. 2(b). Good agreement is achieved. Plasma density decreases exponentially along the radial direction, at the axis the plasma density is of the order of $10^{22}$ m$^{-3}$.

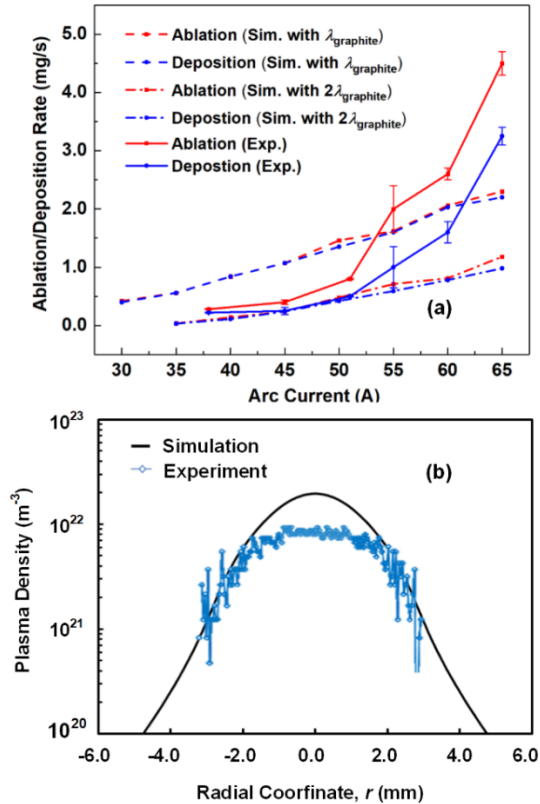

**Fig. 2** (a) Ablation/deposition rate as a function of arc current obtained from the simulations and the experiments ($d$=1.5 mm). Data of the two series of simulations with different electrode thermal conductivities, $\lambda_{graphite}$ (taken from Ref. [40]) and $2\lambda_{graphite}$, are plotted with dash and dash-dot lines, respectively. Experimental data are plotted with solid lines; (b) Profiles of plasma number density at the mid-plane between two electrodes obtained from the simulation and experiment ($d$=3.0 mm and $I$=55 A).

## B. 2D modeling results of carbon arc discharge
### 1. *Profiles of plasma properties and potential along the axis*

Profiles of the plasma number density and space potential along the axis are presented in Fig. 3. Note that no a priori equilibrium assumption is adopted in the simulations. Apparently, the local chemical equilibrium can be achieved in the arc column, where the plasma number density is very close to the equilibrium number density obtained from the Saha equation. The electric field is relatively small and nearly constant in this region. At the vicinity of electrodes, chemical equilibrium breaks down and the number density gradient dramatically increases. The electric field changes the sign near the anode due to high-density gradients there. Cathode sheath voltage drop is negative as expected with the voltage drop of the order of 6V, but the anode sheath is always positive with the voltage drop of the order of 1V. We only observed positive anode sheath in carbon arc in agreement with our earlier study of the argon arc[27] where positive sheath voltage was observed unless the anode was artificially maintained cool. Recent experimental measurements of carbon arc discharges at Princeton Plasma Physics Laboratory (PPPL) also report positive anode sheath[45]. The reason for the positive sheath voltage formation is the necessity to limit high electron emission current from the anode surface as compared to the arc current.

We also disclose that the measured anode voltage drop[45] is several volts higher than the simulated value. We believe that this is because large micron-sized carbon dust particles are injected from the anode[46]. Carbon particles provide the effective recombination surface for the plasma and therefore can greatly increase the plasma resistivity near the anode, therefore leading to a larger anode voltage drop in experiments.

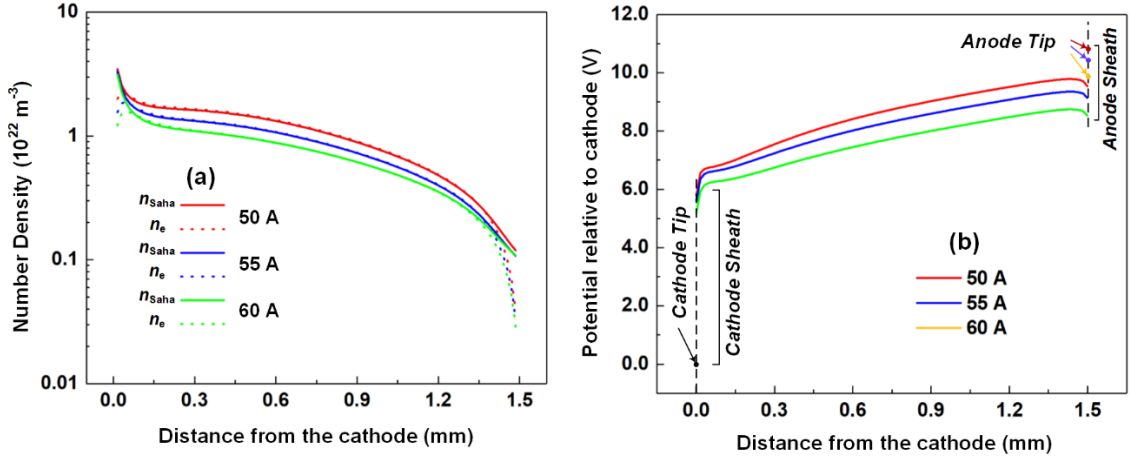

**Fig. 3** Profiles of plasma number density (a) and potential (b) along the axis for the cases with $d$=1.5 mm and $I$=50 A, 55 A and 60 A. In Fig. 3(a), $n_e$ and $n_{Saha}$ represent the simulated electron number density and equilibrium electron number density calculated using the Saha equation, respectively.

Profiles of current density and temperature of electrons and heavy particles along the axis are plotted in Fig. 4. Magnitude of the current density decreases with distance to the cathode surface. For all cases presented in this paper, arc gap is too short to reach local thermal equilibrium ($T<T_e$ throughout the arc). Both the cathode and anode temperatures are very close to the carbon vapor saturation temperature, 3834 K. This is because radiation heat exchange establishes near-equilibrium between both electrodes.

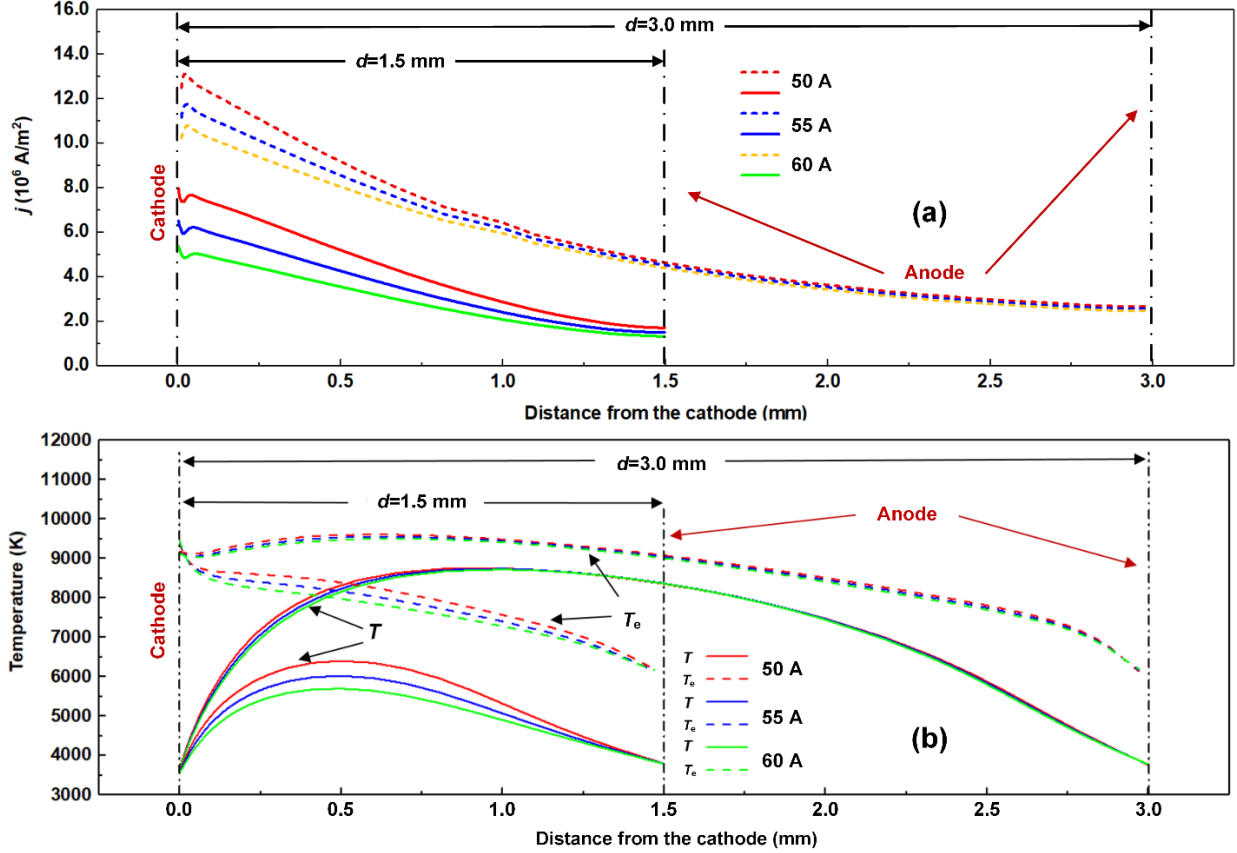

**Fig. 4** Current density profiles (a) and temperature profiles of heavy particles and electrons (b) along the axis for the cases with $d$=1.5 mm and 3.0 mm, and $I$=50 A, 55 A and 60 A.

## 2. *Formation of anode and cathode spots.*

Radial current density profiles on the cathode and anode front surfaces for the cases with $d$=1.5 mm and 3.0 mm are plotted in Fig. 5. The magnitude of the current density inside the spots reaches $10^6$ A/m$^2$. *Interestingly, the anode spot radius increases with the arc current, eventually occupying the whole anode*. Local current density in the anode spot is 2~4 times lower than that in the cathode spot. Formation of the spots indicates that most of the current flows through the path between the two spots and the entire arc plasma becomes constricted. Contours of other plasma parameters and current streamlines are given in Appendix D.

Temperature profiles on the anode and cathode tips for the cases with $d$=1.5 mm and 3.0 mm are shown in Fig. 6. Both the anode and cathode surface temperature exhibit the flat-top profiles, forming the temperature spot. Current and temperature profiles are strongly correlated due to the strong dependence of the local current density on the local temperature due to ablation and emission processes. Temperature inside the anode spot is close to the carbon vapor saturation temperature, $T_{\text{sat, 68kPa}}$=3834 K, whereas the temperature in the cathode spots is a bit lower.

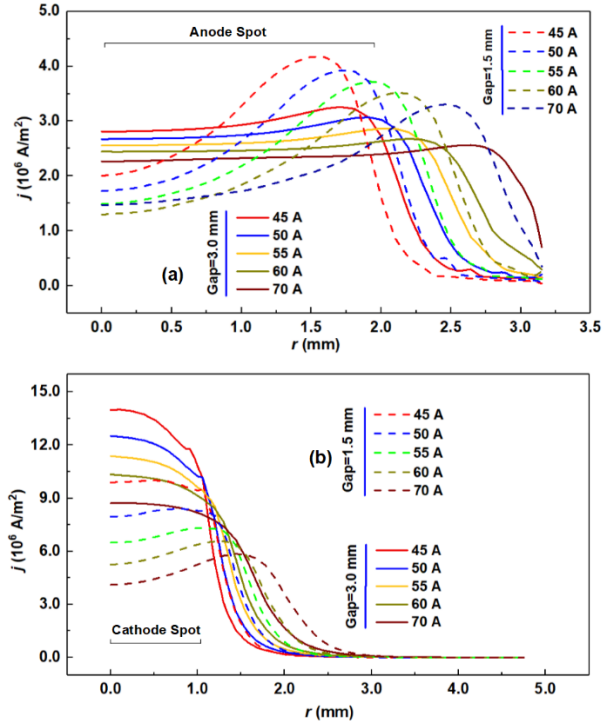

**Fig. 5** Current density profiles along the anode (a) and cathode (b) surface as a function of radius. Cases shown are for: $d$=1.5 mm (1) and $d$=3.0 mm (2), and current $I$= 45 A, 50 A, 55 A, 60 A and 70 A.

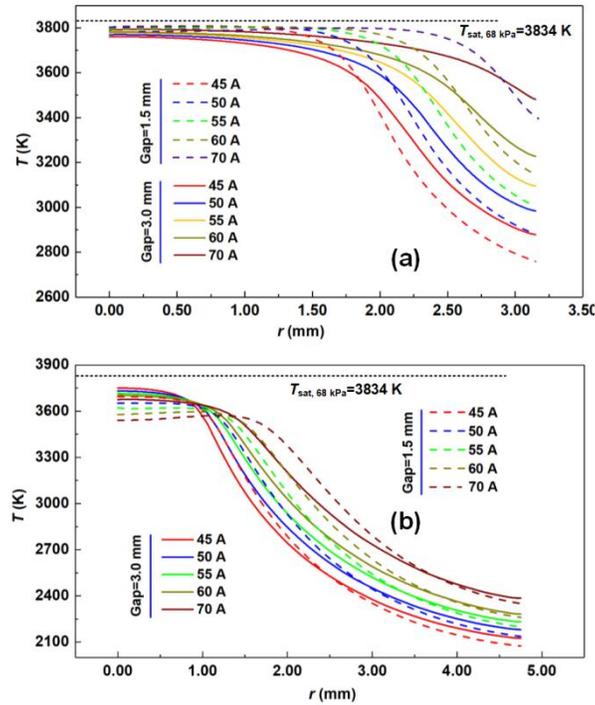

**Fig. 6** Temperature profiles along the anode (a) and cathode (b) surface as a function of radius for two gap widths: $d$=1.5 mm (1) and $d$=3.0 mm (2), and 5 values of current $I$= 45 A, 50 A, 55 A, 60 A and 70 A.

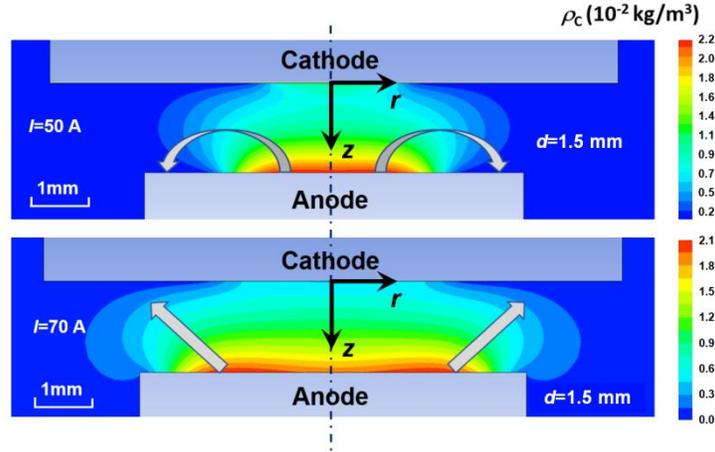

**Fig. 7** Contours of carbon gas density ($d$=1.5 mm, $I$=50 A and 70 A).

## 3. Effects of the spot formation on carbon gas transport and anode ablation

It is commonly believed that the ablated carbon gas from the anode surface can move both to the cathode and out of the gap between the two electrodes. However, it is found that an anomalous carbon gas transport process is established due to the existence of anode spot, and therefore a fraction of ablated carbon deposits back to the anode. Contours of carbon gas density for the cases with $d$=1.5 mm and profiles of ablation rate on the anode surface are illustrated in Fig. 7 and Fig. 8. As already shown in Fig. 6, the surface temperature in the anode spot is very close to the temperature of carbon evaporation. Because of high carbon ablation, at the vicinity of anode spot, nearly all of helium is substituted by the ablating carbon. As shown in Fig. 7, carbon gas density is very high in the spot, but whereas dramatically decreases in the periphery due to the decrease of the surface temperature. The large carbon gas density gradient is, consequently, created and drives the carbon particles ablating from the anode spot towards the periphery and back to anode surface where it deposits due to lower temperature at the periphery (see Fig. 8). A fraction of carbon that returns to the anode at its periphery decreases with the total arc current. For example, for arc current, $I$=45 A, 35% of the ablating carbon deposits on the periphery of anode, whereas for $I$=70 A, only 14% deposits back. Returning of the ablated carbon reduces the total ablation rate and can reduce production of the nanoparticles. Note that in a real arc, spot moves along the surface from the center to periphery and this leads to averaging of the ablation rate over the anode surface.

The flow pattern is shown in Fig. 9. For $I$=60 A, the simulation shows that carbon evaporation produces a maximum velocity of 9.5 m/s, which is ten times lower than the value (~95 m/s) obtained in Ref. [29]. The difference in the flow velocity is due to the difference in the ablation rate relation used in two models. In Ref. [29], the partial pressure of carbon gas was not taken into account for calculating the ablation flux [compare the Eq. (7) in Ref. 29 and the Eq. (17) in Sec. II]. The flow pattern also evidences the anomalous transport of ablated carbon gas to the anode periphery. More details of the effects that the carbon gas transport has on the electrode ablation and deposition can be found in our separated analytical paper, Ref. [47].

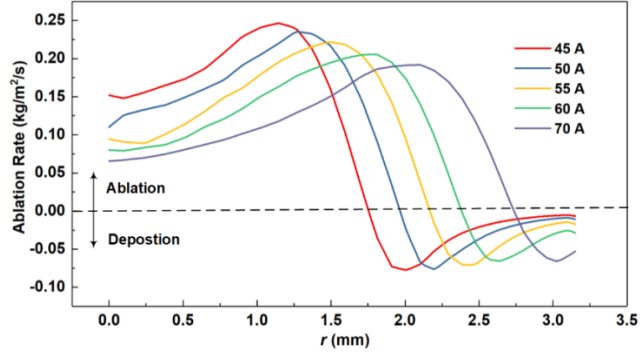

**Fig. 8** Ablation rate at the anode tip. Gap width is fixed as 1.5 mm, and arc current is varied: 45 A, 50 A, 55 A, 60 A and 70 A.

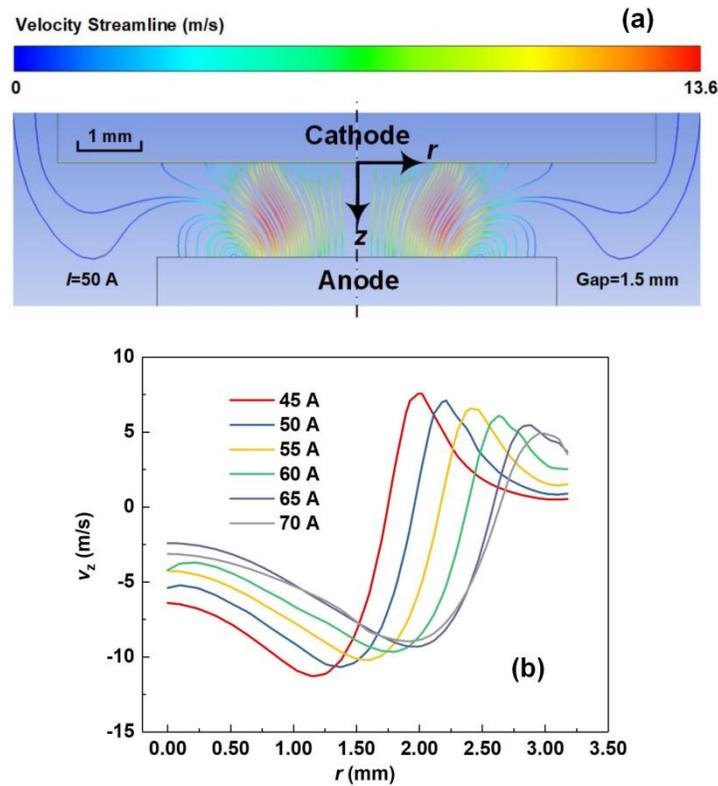

**Fig. 9** (a) Velocity streamlines ($d$=1.5 mm, $I$=50 A); (b) profiles of carbon gas axial velocity ($v_z$) along the anode front surface ($d$=1.5 mm, $I$=45 A, 50 A, 55 A, 60 A, 65 A and 70 A).

### C. Mechanism of the anode spot formation and its radius estimate

#### 1. *Experimental observation of the anode spot in carbon arc discharges*

The formation of anode spot has already been observed in carbon arc discharges, see e.g., Ref. [9]. It was observed that the spot size depends on the total arc current and remains almost the same despite its movement along the anode surface. Vlad Vekselman performed similar experiments at Princeton Plasma Physics Laboratory. A detailed arc synthesis setup has been introduced previously in Ref. [35] and therefore is not described in this paper. For better observation of the anode surface, anode rod was cut at a 45-degree angle with respect to the arc axis and CCD cameras were placed facing the setup in the direction perpendicular to the axis. Figure 10 shows several snapshots for *I*=50 A and 60 A. Anode spot (identified as the yellow region in Fig. 10) moves over the anode surface. However instantaneous arc size is approximately the same for given arc current. This movement cannot be captured in our steady-state axisymmetric simulations. But our simulations can successfully reproduce anode spot structures and their size. This provides a good motivation to identify the anode spot formation mechanism.

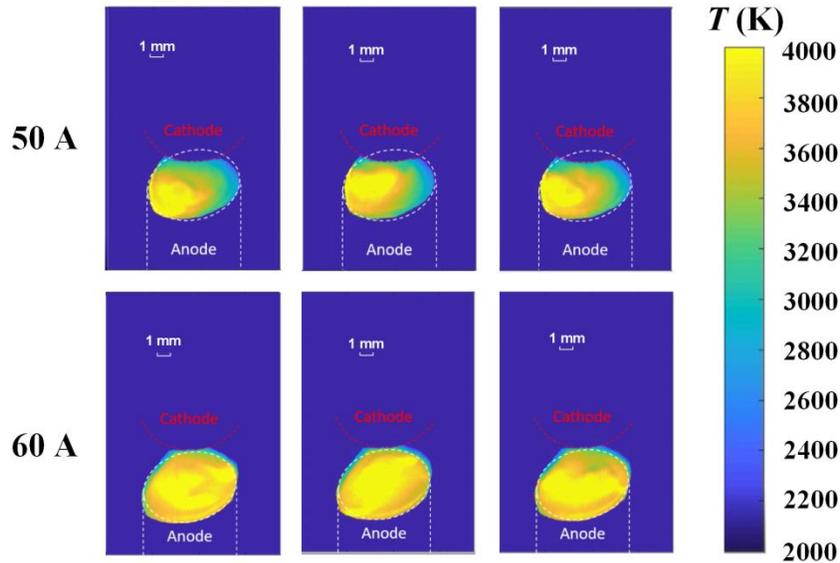

**Fig. 10** Experimental observation of anode spot by V. Vekselman in the setup described by Ref. [35].

## 2. Mechanisms of the anode spot formation

Typically, anode spot formation is attributed to evaporation-ionization or thermal instability, see review [48]. In contrast, here, we proposed a novel idea that the nonlinear nature of the heat balance in the anode body is the main mechanism for the anode spot formation.

In order to identify the mechanism, we performed several series of simulations where different properties were varied. In the first series of simulations, we varied the plasma transport coefficients, e.g., $Q^{e\text{-}h}$ and $D_a$. In the second series of simulations, we varied the electrode properties, e.g., $\lambda_{graphite}$, $\varepsilon$, and $T_{sat,\ 68kPa}$. The case with *d*=3.0 mm and *I*=50 A is chosen as the reference case. Current density profiles on the anode tip for different cases are plotted in Fig. 11. As apparent from the figure, variation of the plasma transport coefficients can hardly affect the spot radius, whereas the spot size is strongly affected by variation of the anode properties. This indicates that the transport processes in the anode body, not in the arc plasma, govern the anode spot formation.

Based on this observation, we propose that the spot formation is the result of complex nonlinear heat balance in the anode. The heat flux from plasma and incident radiation from the cathode are balanced by the heat conduction through the anode body and heat losses from its surface due to electron emission, radiation and ablation. In our cases, ablation rate is not so high, and emissive current is limited by positive sheath voltage. Hence, most of the incoming heat power is balanced by the radiation loss (see Appendix E for more information). For such a cylindrical heat conduction problem with the radiation boundary condition, normally the surface temperature on the anode front surface decreases monotonically with the radius. In addition, due to very strong ablation flux which would be for temperatures above the carbon gas saturation temperature, the maximum of the surface temperature cannot exceed the saturation temperature, $T_{sat,68kPa}$. Therefore, the temperature profile forms a nearly flat-top distribution resembling an anode spot, as shown in Fig.11.

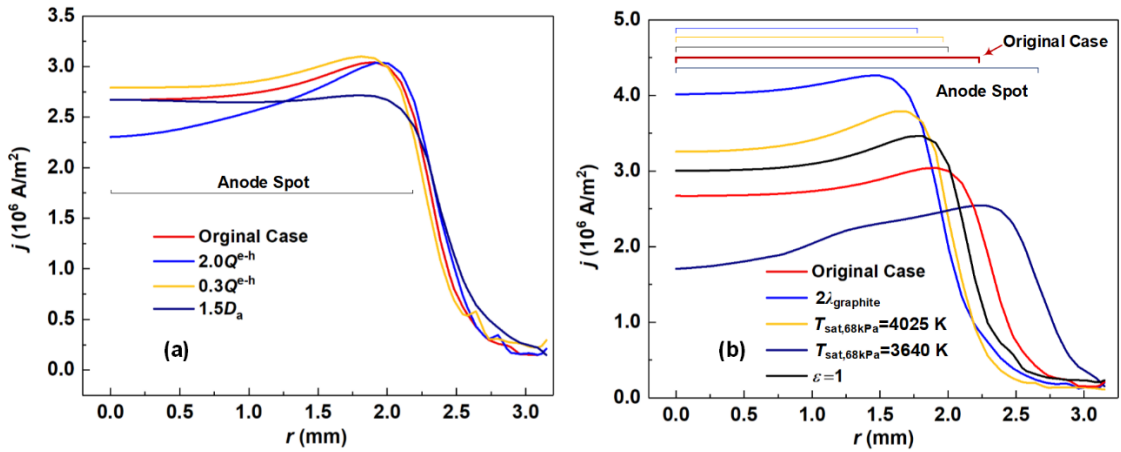

**Fig. 11** Current density distributions over the anode tip with altering the plasma transport (a) and anode properties (b). $d$=3.0 mm and $I$=50 A.

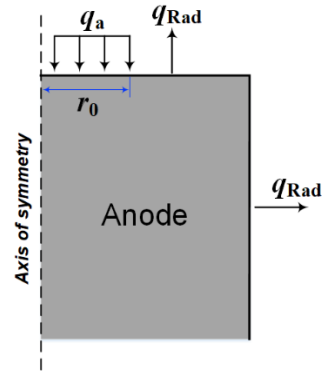

**Fig. 12** Schematic of the reduced simulation setup.

In order to prove this hypothesis, a reduced for the heat balance in the anode body was developed and implemented in ANSYS-CFX to predict the anode spot radius. For the sake of a specific definition of spot radius, we will define the anode spot as the region with the temperature above 3668 K. At surface temperature 3668 K, the carbon saturated vapor pressure is 1/e times of the reference background gas pressure in the chamber [$p_{Sat}$(3668 K)= $p_{Sat}$(3834 K)/e=25 kPa]. For the cases with not too small arc gap width or not too high arc current, the absorbed radiation from the cathode is usually of minor significance and, therefore, is neglected in the reduced model. According to Ref. 49, the total net power from the plasma to the anode is

$$Q_{anode} = \left[ \frac{2.5kT_{e,anode}}{e} + V_w + \max(V_{sheath}, 0) \right] I, \qquad (27)$$

where $T_{e,anode}$ and $V_{sheath}$ are typically of the order of 6000 K and 1 V as obtained in our simulations, respectively, and the work function of graphite, $V_w$, is 4.6 V. Therefore, the total heat flux to the anode is

$$Q_{anode} \approx I \cdot 6.9 [V]. \qquad (28)$$

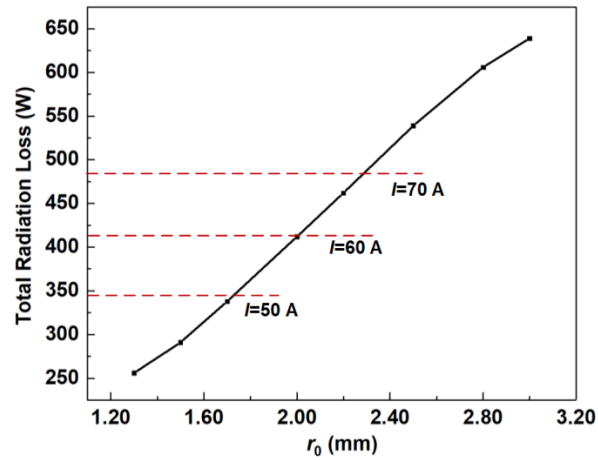

**Fig. 13** Radiation loss versus spot radius obtained from the reduced simulations (Black solid line). Red dashed lines represent the total heat flux to the anode calculated by Eq. (28).

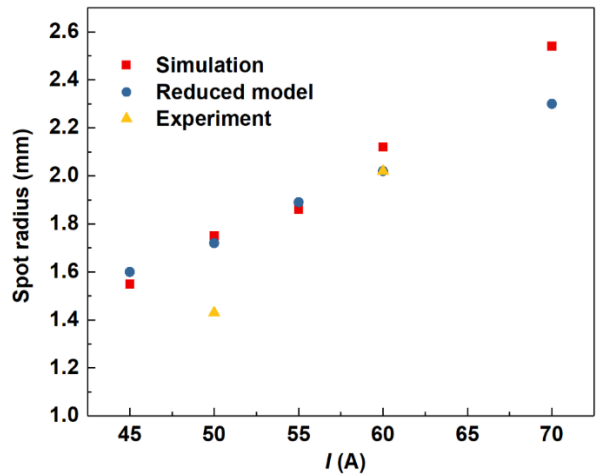

**Fig. 14** Comparisons of the simulated and estimated spot radiuses versus the total arc current.

Radiation loss through the anode surface can be obtained making use of ANSYS-CFX simulations of the heat conduction process in the cylindrical anode body. The schematic of the simulation setup is shown in Fig. 12. Anode Joule heating is proved to have little effect on the temperature profile in the anode and thus can be omitted[49]. Temperature in the anode spot is fixed at the assumed value, 3668 K, see Fig.6. In order to maintain a fixed temperature in the spot, local boundary heat flux is automatically determined within the anode spot and is set to zero outside, i.e.,

$$q_{\text{in, anode tip}} = \begin{cases} q_a, & r < r_0 \\ 0, & r > r_0 \end{cases}, \tag{29}$$

where $r_0$ is the radius of the anode spot (an input parameter for the simulation). The incoming heat flux is balanced by radiation loss, $q_{\text{Rad}} = \varepsilon \sigma_{\text{SB}} T^4$, from all the anode surfaces. Note that $q_a$ here is not determined by $Q_{\text{anode}}$ in Eq. (27).

Fig. 13 maps the total radiation loss versus spot radius. Red dash lines show the total incoming power ($Q_{\text{anode}}$) and corresponding arc current [Eq. (28)]. Spot radius can be determined by matching the incoming power with the total radiation loss (the intersection of red lines and black curve).

Experimental data in Fig. 10 is also processed to obtain the measured spot radius. We can define the measured spot radius as $r_{0,\text{Exp}} = \sqrt{\dfrac{S_{\text{Spot}}}{\pi}}$, where $S_{\text{spot}}$ represents the total size of regions where the temperature is above the threshold temperature, 3668 K. Spot radii obtained from the reduced model, the full simulation (for the case with $d$=3.0 mm) and experiments are plotted in Fig. 14. Relatively good agreement is achieved proving our hypothesis that the heat balance in the anode plays the primary role in the anode spot formation, with a major role of heat flux from the plasma and surface radiation. Increasing differences in the results of the full and reduced models with arc current can be attributed to absorbed radiation from the cathode not included in the reduced model.

## IV. SUMMARY

In this paper, two-dimensional simulations of short carbon arc discharges are performed with a self-consistent fluid model. Entire arc, electrodes and chamber are included in the model for studying heat and particle transport in the arc. Space-charge limited sheaths are considered as the boundary conditions at electrode surfaces. Ablation/deposition, electron emission, radiation and ionization/recombination are included in the heat transfer model. The arc model was implemented into a robust commercial software ANSYS-CFX which was highly customized for this purpose. The model was validated by comparing it to available experimental data. Good agreement on the plasma density profile at the arc midplane and the anode spot size was obtained.

Based on the simulation results, the following conclusions are drawn.

(1) Boundary conditions at the plasma-electrode interfaces strongly affect plasma and gas properties: the current density and heat flux are affected by the sheath boundary conditions, and the carbon density is determined by its ablation and deposition from the electrode surfaces.

(2) Carbon arc reaches local chemical equilibrium (LCE) to satisfy the Saha equation in the arc column, but the local thermal equilibrium (LTE) is not achieved throughout the entire arc, i.e. the electron temperature is always higher than the gas temperature. For typical cases corresponding to the experimental conditions, negative anode sheath voltage drop is never observed in the simulations.

(3) Spot formation is observed on the front surfaces of both the cathode and anode, where the current density, surface temperature, and ablation rate are at their highest. Simulations and a reduced heat-transfer model show that the anode spot radius increases with the arc current in accordance with experimental observations. Due to the anode spot formation, some of the ablated carbon from the spot region returns back to anode periphery, thereby reducing the total ablation rate.

(4) A mechanism of anode spot formation is analyzed. It is proposed that heat balance in the anode and a cutoff of the temperature profile by the carbon vapor saturation temperature govern the anode spot formation.

(5) The reduced heat transfer model was utilized to estimate the anode spot radius based on the heat transfer in the anode body with a balance of radiation loss and heat flux from the plasma at the anode surface. Spot radii obtained from the reduced model are in good agreement with those determined from full simulations and experiments. This further proves the hypothesis that the anode spot formation is governed by nonlinear heat transport in the anode body, with a major role of heat flux from the plasma and surface radiation.

(6) Anode ablation rate and spot size are very sensitive to the anode properties, in particular, thermal conductivity, suggesting more experiments to obtain reliable material properties data.

## Acknowledgments

The work of Jian Chen was supported by the China Scholarship Council and this research was funded by the US Department of Energy Fusion Energy Sciences. Authors are very grateful to Dr. Yevgeny Raitses, Dr. Alexander V. Khrabrov, Dr. Shurik Yatom and Nirbhav S. Chopra for fruitful discussions and Dr. Valerian Nemchinsky for his advice in developing the arc model.

## Data Availability Statement

The data that support the findings of this study are available from the corresponding author upon reasonable request.

## APPENDIX A: Ambipolar approximation of the ion transport equation.

For electrons and singly charged carbon ions, total fluxes are

$$\vec{\Gamma}_e = -\mu_e n_e \vec{E} - D_e \nabla n_e - D_{Te} \nabla T_e + n_e \vec{v}$$
$$\vec{\Gamma}_i = \mu_i n_i \vec{E} - D_i \nabla n_i - D_T \nabla T + n_i \vec{v}$$
(A1)

where the diffusion coefficients are $D_e = \dfrac{\mu_e k T_e}{e}$, $D_i = \dfrac{\mu_i k T_i}{e}$, $D_{Te} = \dfrac{\mu_e k n_e}{e}$, $D_T = \dfrac{\mu_i k n_i}{e}$. Given that $n_e = n_i$ and $\vec{\Gamma}_i - \vec{\Gamma}_e = \dfrac{\vec{j}}{e}$, divide $\Gamma_e$ by $\mu_e$ and $\Gamma_i$ by $\mu_i$, and write their sum as

$$\frac{\vec{\Gamma}_i}{\mu_e} + \frac{\vec{\Gamma}_i}{\mu_i} - \frac{\vec{j}}{e\mu_e} = -\left(\frac{D_e}{\mu_e} + \frac{D_i}{\mu_i}\right)\nabla n_i - \frac{D_{Te}}{\mu_e}\nabla T_e - \frac{D_T}{\mu_i}\nabla T + \left(\frac{n_i \vec{v}}{\mu_e} + \frac{n_i \vec{v}}{\mu_i}\right).$$
(A2)

Provided that $\nabla \cdot \vec{j} = 0$ and neglecting spatial variation of all the transport coefficients, Eq. (A2) can be transformed to

$$\frac{\mu_e + \mu_i}{\mu_e \mu_i} \nabla \cdot \vec{\Gamma}_i = -\nabla \cdot \left(\frac{\mu_e D_i + \mu_i D_e}{\mu_e \mu_i}\nabla n_i\right) + \frac{\mu_e + \mu_i}{\mu_e \mu_i}\nabla \cdot (n_i \vec{v}) - \nabla \cdot \left(\frac{D_{Te}}{\mu_e}\nabla T_e + \frac{D_T}{\mu_i}\nabla T\right).$$
(A3)

And the continuity equation for ions is $\nabla \cdot \vec{\Gamma}_i = S_i$, and hence,

$$\nabla \cdot (n_i \vec{v}) = \nabla \cdot \left(\frac{\mu_e D_i + \mu_i D_e}{\mu_e + \mu_i}\nabla n_i\right) + S_i + \nabla \cdot \left(\frac{\mu_i D_{Te}}{\mu_e + \mu_i}\nabla T_e + \frac{\mu_e D_T}{\mu_e + \mu_i}\nabla T\right).$$
(A4)

Here, the third term (thermal diffusion term) at the right hand of Eq. (A4) is small compared to the first term [$\dfrac{\nabla T_e}{T_e}$ and $\dfrac{\nabla T}{T} \ll \dfrac{\nabla n_i}{n_i}$, see Fig. 3(a) and Fig. 4(b)] and can be neglected. Neglecting this term yields:

$$\nabla \cdot (n_i \vec{v}) = \nabla \cdot (D_a \nabla n_i) + S_i.$$
(A5)

Note that solving ion diffusion in the form (A5) implies a modification of boundary conditions for ion flux. Ambipolar flux $\vec{\Gamma}_a = -D_a \nabla n_i$ is not physically equal to the actual ion flux, but $\vec{\Gamma}_i = \mu_i n_i \vec{E} - D_i \nabla n_i - D_T \nabla T + n_i \vec{v}$. According to Eq. (A2) and neglect the minor terms, boundary condition on the ion flux should be given as

$$D_a \nabla n_i = -\frac{\vec{j}_i}{e} + \frac{\mu_i}{\mu_e + \mu_i}\frac{\vec{j}}{e}.$$
(A6)

### APPENDIX B: Insignificance of the net radiative loss from the bulk plasma.

This appendix show the reasons why the radiative loss from arc plasma is omitted in our modeling. On one hand, experiment measurements showed that the radiation from arc plasma is much smaller compared to the radiation from the two electrodes [41]. One the other hand, the simple estimation using the net emission approximation [50] also proves that this term is insignificant. It is noted that although the net emission approximation is not accurate for the strong non-equilibrium cases, the net emission coefficient, $\varepsilon_N$, can be used as the first estimation of the net radiative loss from plasma[50]. As shown in Fig. 5 of Ref. [51], the calculated $\varepsilon_N$ of carbon plasma with a mean plasma radius 2 mm at atmospheric pressure varies from $2\times10^7$ W/m$^3$/sr to $10^8$ W/m$^3$/sr at the temperature range [5 kK, 10 kK] that covers the typical maximum electron and heavy particle temperature in carbon arcs. Therefore, the volumetric net radiative loss term, $q_{r,N}=4\pi\varepsilon_N$, varies from $2.5\times10^8$ W/m$^3$ to $1.3\times10^9$ W/m$^3$.

Profiles of the energy source terms in electron energy transport equation [Eq. (12)] along the axis for arcs with $d=1.5$ mm and $I=50$ A are plotted in Fig. 15. As seen, the heat exchange and Joule heating are dominant in the bulk plasma, which are of the order of magnitude $10^{10}$ W/m$^3$. $q_{r,N}$ is typically several ten times less than the heat exchange term, which means that the radiative loss in the bulk plasma for typical carbon arcs are not significant for the energy transport of both electron and heavy particles. Thus, radiative loss from the bulk plasma can be omitted in our simulations. Note that the radiative loss from the bulk plasma is also neglected in recent modeling studies of carbon arcs[28, 29].

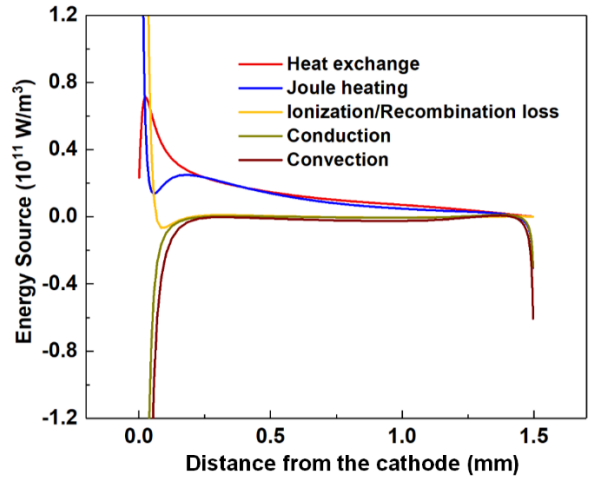

Fig. 15 Profiles of the energy source terms in electron energy transport equation [Eq. (12)] along the axis for arcs with $d=1.5$ mm and $I=50$ A.

## APPENDIX C: Sufficiency of the computational grid resolution check.

In arc discharges, the gradients near the solid-fluid interface are much larger than those in the plasma. Some of them, such as the charge density gradient, play an important role in conducting the current[18]. Thus, computational meshes were refined at the vicinity of the fluid-solid interface to guarantee a sufficient resolution (see Fig. 16).

For the arc modeling in this paper, 36252 grid elements for arcs with $d=1.5$ mm and 36732 grid elements for $d=3.0$ mm were used. Check of the grid resolution sufficiency is performed for a typical case with $d=1.5$ mm and $I=60$ A by doubling the element number in both radial and axial dimensions. Profiles of heavy particle and electron temperature along the axis with the refined mesh (145008 elements) and the present mesh (36252 elements) are plotted in Fig. 17(a) and 17(b), respectively. Maximum deviations between the two cases are 8% for the heavy particle temperature and 3% for the electron temperature showing weak solution dependence on the grid resolution.

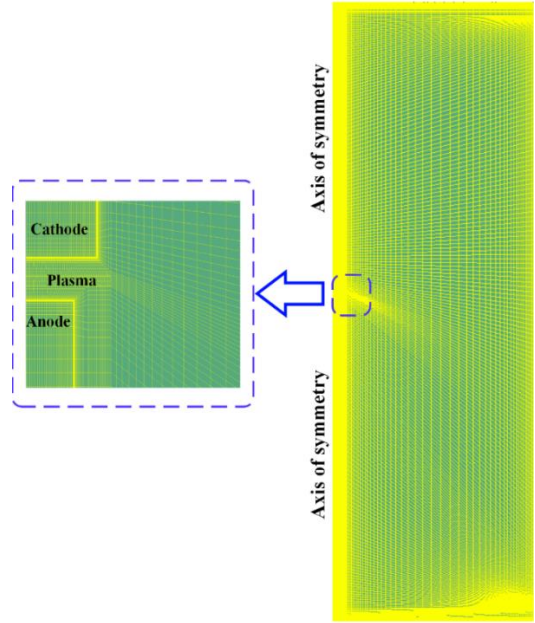

**Fig. 16** Computational grid used for the cases with *d*=1.5 mm.

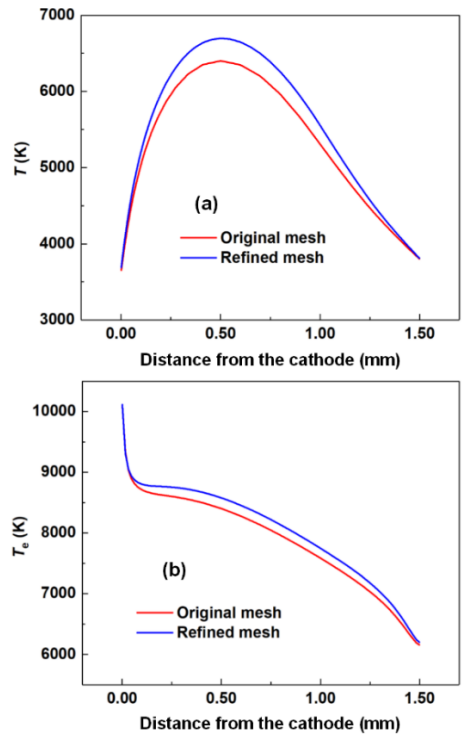

**Fig. 17** Comparison of the heavy particle temperature (a) and electron temperature (b) along the axis (*r*=0). Results are obtained with the original (red) and refined (blue) meshes, respectively. The refined mesh used a double element number than the original mesh in both radial and axial dimensions.

## APPENDIX D: Contours of parameters and current streamlines.

Contours of $T$, $T_e$, $n_e$ and $\varphi$ for the cases with $I$=50 A, $d$=1.5 mm and 3.0 mm are presented in Fig. 18. The heavy particle temperature is high near the cathode, forming the arc core. Gradients of $T$, $T_e$, $n_e$, $\varphi$ show a strong correlation with the current density. The entire arc is constricted by the two spots and does not reach LTE state. Plasma density is at the order of $10^{22}$ m$^{-3}$. For the cases with $d$=1.5 mm, the highest potential is located not at the axis but a bit away because the maximum current density is there.

Current streamlines for the case with $d$=3.0 mm and $I$=50 A are plotted in Fig. 19. It is seen that the current density near the spots is very high. Most of the current flows from the anode spot to cathode spot, and therefore the arc is constricted.

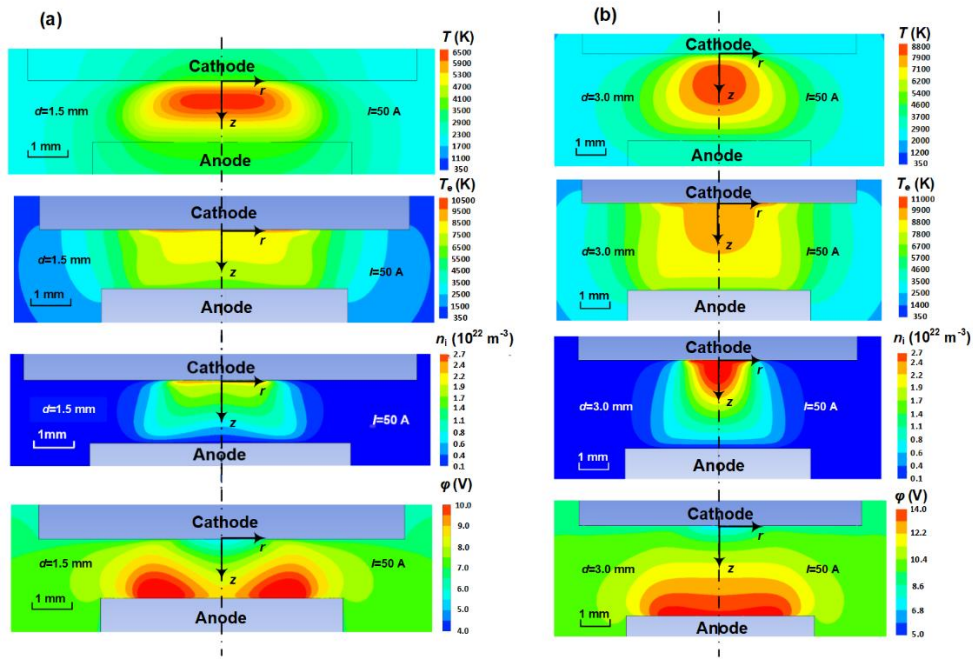

**Fig. 18** Contour of the temperature, electron temperature, plasma number density and potential ($d$=1.5 mm and $d$=3.0 mm, $I$=50 A).

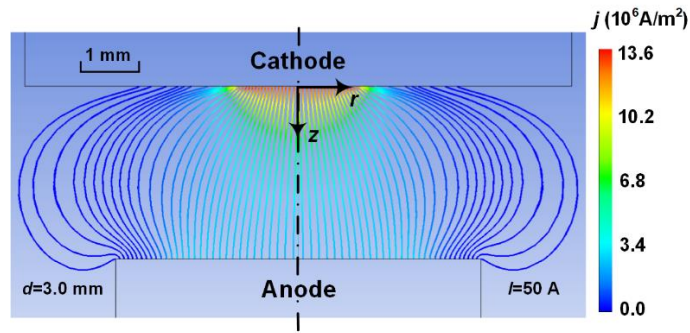

**Fig. 19** Current streamlines for the case with $d$=3.0 mm and $I$=50 A.

**APPENDIX E: Analysis of heat flux components on the anode and cathode tips.**

In this appendix, the power balance on the anode and cathode tips is analyzed. As presented in Fig. 20, we categorized the power transferred from the environment to the electrodes as the power input (including absorbed radiation, conductive heat from the plasma, plasma electron heat, ion heat and carbon deposition heat), and the power from the electrodes to the environment as a power loss (including ablation energy loss, conductive heat into electrode body, emitted electron energy loss and radiation loss from the electrode tip). Note that the conductive heat into the electrode body is categorized as power loss because it finally dissipates through the radiation from the electrode surfaces.

Here, components of total heat flux across the tips for the cases with $d$=1.5 mm and different arc currents are illustrated in Fig. 21. At the anode tip, the surface temperature is sustained mainly by the heat from plasma. Ion heat is small, because the anode sheath voltage drop is positive. Absorbed radiation and conductive heat from plasma are several times less than the plasma electron heat flux introduced by electrons absorbed by the surface. As for the power loss, radiation loss from the anode tip and conductive heat into the anode body dominate. Since the conductive heat flux into the anode body eventually dissipates by the radiation from the side surfaces, the power input is mostly converted into the radiation loss from the whole anode surface. At the cathode tip, ion heat, conductive heat from the plasma and absorbed radiation constitute comparable contributions. Ref. [52] provided an analytical expression to estimate the absorbed heat across the cathode tip

$$P_A = F_{ac}\varepsilon^2 \sigma_{SB} T^4,$$
$$F_{ac} = \frac{1}{2}\left( X - \sqrt{X^2 - 4\left(\frac{r_c^2}{r_a^2}\right)} \right),$$
(E1)

where $F_{ac}$ is the view factor, $\sigma_{SB}$=5.67×10$^{-8}$ W/m$^2$/K$^4$ is the Stephan-Boltzmann constant, the radii of cathode and anode are $r_c$ and $r_a$, respectively, $X = 1 + \frac{d^2}{r_a^2} + \frac{r_c^2}{r_a^2}$. Note that uniform anode surface temperature is assumed when using Eq. (E1), which does not correspond to the flat-top temperature distribution over the anode tip. However, some corrections can be made regarding a relatively constant temperature inside the spot. For a rough estimation of the cathode absorbed radiation, we can account for the major contribution which comes from the irradiation of the anode spot. Then, anode radius in Eq. (E1) can be changed to the anode spot radius. The saturation temperature, 3834 K, can be used to approximate the surface temperature inside the spot region. Results from Eq. (E1) are presented in Fig. 21(c). Estimation shows a better agreement with the simulation for the cases with high arc current because the irradiation from the anode periphery is less contributing.

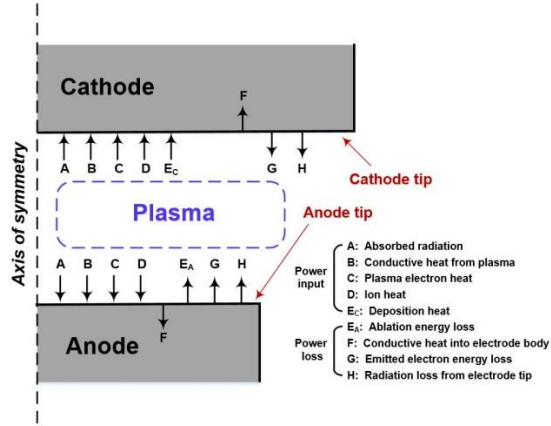

**Fig. 20** Scheme of the power across the electrode front surfaces.

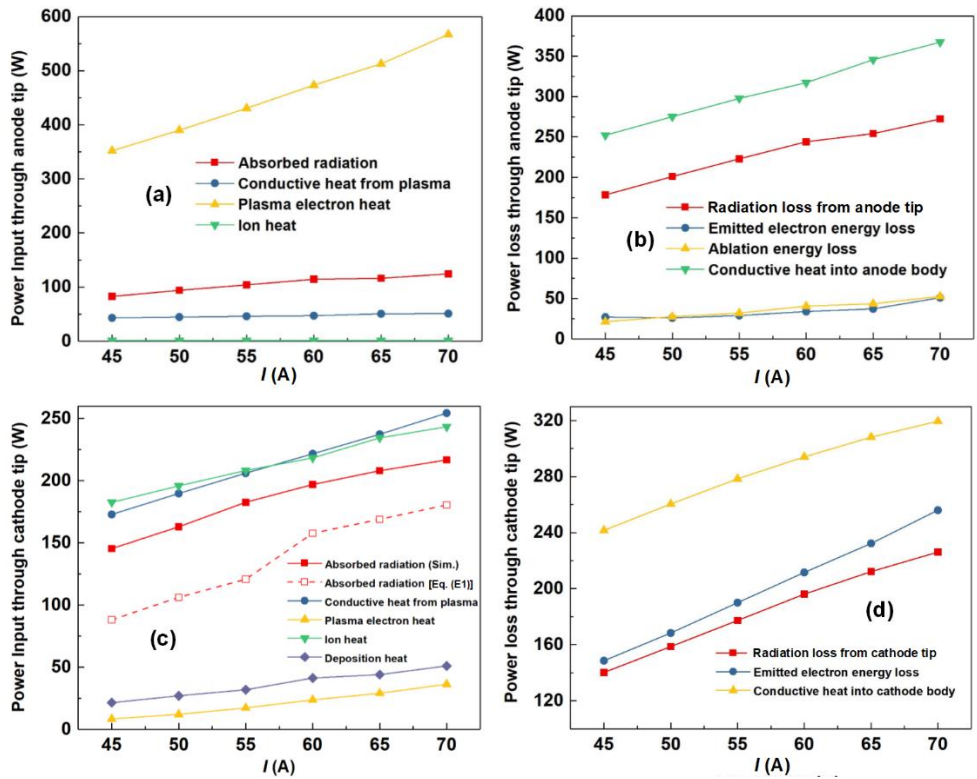

**Fig. 21** Power input and loss through the anode tip [(a), (b)] and the cathode tip [(c), (d)]. Estimation of the absorbed radiation [Dash line in (c)] is derived based on Eq. (E1)[52]